\documentclass[letterpaper, 10 pt, conference]{ieeeconf}  

\IEEEoverridecommandlockouts                              

\usepackage{graphicx}
\usepackage{latexsym}
\usepackage{amsmath}
\usepackage{amssymb}
\usepackage{epsfig}
\usepackage{cite}
\usepackage{algorithmic}
\usepackage{overpic}
\usepackage{multirow}
\usepackage[table]{xcolor}
\usepackage{graphicx}
\usepackage{colortbl,array}
\usepackage{multirow,bigdelim}
\usepackage{graphicx}
\usepackage{subfigure}
\usepackage{booktabs}
\usepackage{booktabs,amsmath}
\usepackage{fancyhdr, graphicx, amsmath, amssymb}
\usepackage[linesnumbered,ruled]{algorithm2e}
\usepackage{arydshln}
\usepackage[english]{babel}
\usepackage{xcolor}
\usepackage{cancel}

\SetCommentSty{mycommfont}

\newtheorem{definition}{Definition}
\newtheorem{prop}{Proposition}
\newtheorem{lem}{Lemma}
\newtheorem{rem}{Remark}
\newtheorem{thm}{Theorem}

\newtheorem{asm}{Assumption}

\usepackage{soul} 

\title{\LARGE \bf
Detectability of Intermittent Zero-Dynamics Attack in Networked Control Systems}

\author{Yanbing~Mao, Hamidreza~Jafarnejadsani, Pan~Zhao, Emrah~Akyol,
        and Naira~Hovakimyan
\thanks{Y.~Mao and E.~Akyol are with the Department of Electrical and Computer Engineering, Binghamton University--SUNY, Binghamton, NY,
13902 USA (e-mail: \{ymao3, eakyol\}@binghamton.edu).}
\thanks{H.~Jafarnejadsani, P.~Zhao and N.~Hovakimyan are with the Department of Mechanical Science and Engineering, University of Illinois at Urbana--Champaign, Urbana, IL, 61801 USA (e-mail: \{jafarne2, panzhao2, nhovakim\}@illinois.edu).}
}

\begin{document}

\maketitle
\thispagestyle{empty}
\pagestyle{empty}

\begin{abstract}
This paper analyzes stealthy attacks, particularly the zero-dynamics attack (ZDA) in networked control systems. ZDA  hides the attack signal in the null-space of the state-space representation of the control system and hence it cannot be detected via conventional detection methods. A natural defense strategy builds on changing the null-space  via switching through a set of topologies. In this paper, we propose a realistic ZDA variation where the attacker is aware of this topology-switching strategy, and hence employs the policy to avoid detection: ``pause (update and resume) attack" before (after) topology switching to evade detection. We first systematically study the proposed ZDA variation, and then develop defense strategies under the realistic assumptions. Particularly, we characterize conditions for detectability of the proposed ZDA variation, in terms of the network topologies to be maintained, the set of agents to be monitored, and the measurements of the monitored agents that should be extracted.  We provide numerical results that demonstrate our theoretical findings.
\end{abstract}

\section{INTRODUCTION}

Security concerns regarding the networked control systems pose significant  challenges to their use and wide deployment, as highlighted by the recent incidents including distributed denial-of-service (DDOS) attack on Estonian web sites~\cite{nazario2009politically}, Maroochy water breach \cite{slay2007lessons} and  cyber attacks on smart grids \cite{cnn2}.  Particularly,  a special class of "stealthy" attacks, namely the  ``zero-dynamics attack" (ZDA), poses a formidable security challenge~\cite{MP15,hl18,MP18}. The main idea behind the machinery of ZDA is to hide the attack signal in the null-space of the state-space representation of the networked control system so that the attack can evade conventional detection methods that are based on the observation signal (hence, the name ``stealthy" attack).  The objective of such a stealthy attack can vary from manipulating the controller to accept false data which would yield the system towards a desired (e.g., unstable) state to altering the system trajectory by maliciously changing the system dynamics (e.g., a stealthy topology attack).

Recent research efforts in this area have mainly focused on two directions: i) novel ZDA variations for particular systems \cite{park2016}, and ii)  defense strategies \cite{n11,n12,F13,weerakkody2017robust,chen2017protecting}.  For example, Park et al.~\cite{park2016} designed a robust ZDA for  a particular stochastic cyber-physical system, for which  the attack-detection signal can be guaranteed to stay below a threshold over a finite horizon. Jafarnejadsani et al.~\cite{jafarnejadsani2017dual,hl18} proposed a multi-rate $\mathcal{L}_{1}$ adaptive controller which can detect ZDA in the sampled-data control systems, by removing certain unstable zeros of discrete-time systems~\cite{MP15,MP18}.

Most of the prior work on defense strategies for the original ZDA  require limiting assumptions regarding the connectivity of network topology and the number of the misbehaving agents (i.e., the agents under attack) \cite{n11,n12,F13,weerakkody2017robust}.  For example, the detection approach in \cite{chen2017protecting}  works only for the scenario of single misbehaving agent  in second-order systems \cite{F13}, i.e., if there are multiple agents compromised by the attacker, the defense strategy cannot detect ZDA. Teixeira et al.~\cite{T12} showed that the strategic changes in system dynamics could be utilized by defender to detect ZDA, but the approach taken in \cite{T12} requires the attack-starting times to be the initial time and known to the defender. In other words, the defense strategy can fail to work if the defender does not know the attack starting time, as is practically the case for most stealthy attack scenarios.

Towards designing a practical ZDA defense strategy,  strategic topology switching is proposed  in \cite{mao2017strategic,ma1811}. The approach here is based on changing the topology through a carefully crafted set of topologies so that the ZDA can be detected. Controlling topology in this manner  is physically feasible thanks to recent developments in mobile computing, wireless communication and sensing~\cite{hartenstein2008tutorial,mazumder2011wireless}. We note, in passing, that the idea of using the changes in the state-space dynamics to detect ZDA first appeared in \cite{T12}, albeit a realistic mechanism (e.g., changing the system topology) to achieve that objective was only very recently studied in \cite{mao2017strategic,ma1811}. However, the  defense strategy in  \cite{mao2017strategic,ma1811}  still relies on the critical assumption of a naive attacker that does not take the topology switching strategy of the defender into account. 

In this paper, we systematically address the following  problem:  can an attack comprising intermittently pausing, updating and resuming ZDA based on the knowledge of the sequence of topologies that the system goes through, be successful?  We study this attack strategy, which we refer to as {\it intermittent ZDA}, in detail. We next analyze possible optimal defense strategies, beyond switching the topology, against the proposed intermittent ZDA. We  demonstrate our theoretical findings via numerical simulation results.

\section{Preliminaries}
\subsection{Notations}
 We let $\mathbb{R}^{n}$ and $\mathbb{R}^{m \times n}$ denote the set of $\emph{n}$-dimensional real vectors and the set of $m \times n$-dimensional real matrices, respectively. Let $\mathbb{C}$ denote the set of complex numbers, and let $\mathbb{N}$ represent the set of the natural numbers, and $\mathbb{N}_{0}$ = $\mathbb{N}$ $\cup$ $\left\{ 0 \right\}$. Let $\mathbf{1}_{n \times n}$ and $\mathbf{0}_{n \times n}$ be the $n \times n$-dimensional identity matrix and zero matrix, respectively. $\mathbf{1}_{n} \in \mathbb{R}^{n}$ and $\mathbf{0}_{n} \in \mathbb{R}^{n}$ denote the vector with all ones and the vector with all zeros, respectively. The superscript `$\top$' stands for matrix transpose.  $\ker \left( Q \right) \triangleq \left\{ {y: Qy = {\mathbf{0}_n}}, Q \in \mathbb{R}^{n \times n} \right\}$, $A^{-1}\mathbb{F} \triangleq$  $\left\{ {y: Ay \in \mathbb{F}}\right\}$. Also, $\left| \cdot \right|$ denotes the cardinality of a set, or the modulus of a number. ${\mathbb{V}} \backslash \mathbb{K}$ describes the complement set of $\mathbb{K}$ with respect to $\mathbb{V}$. $\lambda_{i}\left( M \right)$ is $i^{\emph{\emph{th}}}$ eigenvalue of matrix $M$. For a matrix $W \in \mathbb{R}^{n \times n}$, ${\left[ {{W}} \right]_{i,j}}$ denotes the element in row $i$ and column $j$.

\subsection{Zero-Dynamics Attack}
Consider the following system, where  $\breve{g}(t)$ represents an  attack signal:
\begin{subequations}
\begin{align}
\dot{\breve{z}}\left( t \right) &= A\breve{z}\left( t \right) + B\breve{g}(t),\\
\breve{y}\left( t \right) &= C\breve{z}\left( t \right) + D\breve{g}(t),
\end{align}\label{eq:sl2}\end{subequations}
and $\breve{z}\left( t \right) \in \mathbb{R}^{\bar{n}}$, $\breve{y}\left( t \right) \in \mathbb{R}^{\bar{m}}$, $\breve{g}(t) \in \mathbb{R}^{\bar{o}}$, $A \in \mathbb{R}^{\bar{n} \times \bar{n}}$,
$B \in \mathbb{R}^{\bar{n} \times \bar{o}}$, $C \in \mathbb{R}^{\bar{m} \times \bar{n}}$,  $D \in \mathbb{R}^{\bar{m} \times \bar{o}}$.

\begin{definition}~\cite{teixeira2012attack}
The attack signal ${\breve{g}}(t) = ge^{\eta t}$ is a ZDA,  if there exist scalar $\eta \in \mathbb{C}$, and nonzero vectors $\mathbf{z}_{0} \in \mathbb{R}^{\bar{n}}$ and ${g} \in \mathbb{R}^{\bar{o}}$, that satisfy
\begin{align}
\left[\!
    \begin{array}{c}
        \mathbf{z}_{0}\\ \hdashline[2pt/2pt]
        { - g}
    \end{array}
\!\right] &\in \ker \left(\left[\!
    \begin{array}{c;{2pt/2pt}c}
        {\eta \mathbf{1}_{\bar{n} \times \bar{n}} - A} \!&\! B\\ \hdashline[2pt/2pt]
        -C \!&\! D
    \end{array}
\!\right]\right).\label{eq:czeo}
\end{align}\label{defzda}
\end{definition}

\section{System Description}
Consider a second-order system  of a population of $n$ agents whose dynamics are governed by the
following equations:
\begin{subequations}\label{eq:MA_sys}
\begin{align}
{\dot x_i}\left( t \right) &= {v_i}\left( t \right),\label{eq:oox1}\\
{{\dot v}_i}\left( t \right) &=  u_{i}(t),\hspace{0.5cm}i = 1, \ldots ,n,\label{eq:oox2}
\end{align}\label{eq:ooiio}\end{subequations}
where $x_{i}(t) \in \mathbb{R}$ is the position, $v_{i}(t) \in \mathbb{R}$ is the velocity, and $u_{i}(t) \in \mathbb{R}$ is the local control input\footnote{Several real-world networked systems that can be represented by~(\ref{eq:ooiio}) including the second-order consensus~\cite{ren2007distributed}, flocking~\cite{olfati2006flocking},  swarming~\cite{gazi2004stability},  velocity synchronization and regulation of relative distances~\cite{tanner2007flocking}, and their applications in the decentralized formation control of mobile robots~\cite{lawton2003decentralized} and spacecrafts~\cite{ren2004decentralized}, and distributed continuous-time optimization~\cite{rahili2017distributed}, etc.}.

The interaction among the agents is modeled by an undirected graph $\mathrm{G} \triangleq (\mathbb{V}, \mathbb{E})$, where
$\mathbb{V}$ $\triangleq$ $\left\{ {1,2, \ldots, n} \right\}$ is the set of vertices that represents the $n$ agents and $\mathbb{E} \subseteq \mathbb{V} \times \mathbb{V}$ is the set of edges of the graph $\mathrm{G}$. The weighted adjacency matrix $\mathcal{A} = \left[ {{a_{ij}}} \right]$ $\in \mathbb{R}^{n \times n}$ of the graph $\mathrm{G}$ is defined as $a_{ij} = a_{ji} > 0$ if $(i, j) \in \mathbb{E}$, and $a_{ij} = a_{ji} = 0$ otherwise. Assume that there are no self-loops, i.e., for any ${i} \in \mathbb{V}$, $a_{ii} = 0$.  The Laplacian matrix of the graph $\mathrm{G}$ is defined as $\mathcal{L} \triangleq \left[ {{l_{ij}}} \right] \in {\mathbb{R}^{n \times n}}$, where ${l_{ii}} \triangleq \sum\limits_{j = 1}^n {{a_{ij}}}$, and ${l_{ij}} \triangleq - {a_{ij}}$ for $i \neq j$.


\begin{definition}
\cite{mei2016distributed} The agents in the system~(\ref{eq:ooiio}) are said to achieve the asymptotic consensus with final zero common velocity if for any initial condition:
\begin{align}
\mathop {\lim }\limits_{t \to \infty } | {{x_i}\left( t \right) \!-\! {x_j}\left( t \right)} | \!=\! 0 \hspace{0.1cm}\emph{\emph{and}} \mathop {\lim }\limits_{t \to \infty } | {{v_i}\left( t \right)} | \!=\! 0, \forall i, j \in \mathbb{V}. \label{eq:defc}
\end{align}
\end{definition}

\subsection{Control Protocol}

In this paper, we will propose a defense strategy that is based on topology switching. We now borrow a control protocol that involves topology switching from~\cite{xie2006consensus,mei2016distributed} to achieve consensus according to~(\ref{eq:defc}) for the agents in system \eqref{eq:ooiio}:
\begin{align}
u_i (t)  =  - v_i (t) + \sum\limits_{j \in \mathbb{V}} {a_{ij}^{\sigma \left( t \right)}\left( {{x_j}\left( t \right) - {x_i}\left( t \right)} \right)}, i \in \mathbb{V}, \label{eq:lci}
\end{align}
where $\sigma (t):[t_{0},\infty ) \to \mathbb{S} \triangleq \{1, \ldots, \mathrm{s}\}$ is the switching signal of the interaction topology of the communication network; $a^{\sigma(t)}_{ij}$ is the entry of the weighted adjacency matrix that describes the activated topology of the communication graph. For a system with finite agents, the topology set $\mathbb{S}$ has finite elements as well.

\subsection{System in the Presence of Attacks}
We let $\mathbb{K}$ $\subseteq$ $\mathbb{V}$ denote the set of misbehaving agents, i.e., the agents under attack. We let the increasingly ordered set $\mathbb{M} \triangleq \left\{ {1,2, \ldots } \right\}$ $\subseteq \mathbb{V}$ denote the set of monitored agents. Under time-dependent switching topology, the multi-agent system in \eqref{eq:ooiio}, with the control input given by \eqref{eq:lci} and the outputs of monitoring agents in $\mathbb{M}$ subject to the ZDA signal $\breve{g}_{i}(t)$, can be written as
\begin{subequations}
\begin{align}
\!\!\!{{\dot{\breve{x}}}_i}\!\left( t \right) &\!=\! {{\breve v}_i}\!\left( t \right) \label{eq:oox1}\\
\!\!\!{{\dot{\breve v}}_i}\!\left( t \right) &\!=\! -{{\breve v}_i}\!\left( t \right) + \!\sum\limits_{i \in \mathbb{V}} \!{a_{ij}^{\sigma \left( t \right)}}\!\!\left( {{{\breve x}_j}\!\left( t \right) \!-\! {{\breve x}_i}\!\left( t \right)} \right)\! + \!\!\left\{ \begin{array}{l}
\hspace{-0.22cm}{\breve{g}}_i\!\left( t \right)\!, i \!\in\! \mathbb{K}\\
\hspace{-0.22cm}0, \hspace{0.1cm} i \!\in\! {\mathbb{V}} \backslash \mathbb{K}
\end{array} \right.\label{eq:oox2}\\
\!\!\!{{\breve{y}}_i}\!\left( t \right) &\!=\! c_{i1}\breve x_{i}(t) + c_{i2}\breve v_{i}(t) + d_i{\breve{g}}_i\!\left( t \right)\!, i \!\in\! \mathbb{M}\label{eq:oox3},
\end{align}\label{eq:oofn}\end{subequations}
where $c_{i1}$'s and $c_{i2}$'s are constant coefficients designed by the defender (system operator), while $d_i$'s are coefficients designed by the attacker.

\begin{rem}
If $d_i \neq 0$ for some $i \in \mathbb{M}$, \eqref{eq:oox3} means that the sensor outputs are under attack. However, the attack form in \eqref{eq:sl2} with the attack policy computation \eqref{eq:czeo} indicate that the ZDA signals for sensor outputs and control inputs are not independent of each other.
\end{rem}

The system in \eqref{eq:oofn} can be equivalently rewritten in the form of a switched system under attack:
\begin{subequations}
\begin{align}
&\dot{{{\breve{z}}}}\left( t \right) = A_{{\sigma}(t)}{{\breve{z}}}\left( t \right) + \breve{g}\left( t\right)  \label{eq:s1a}\\
&\breve{y}\left( t \right) = C\breve{z}(t) + D\breve{g}\left( t\right),\label{eq:s1b}
\end{align}\label{eq:s1}\end{subequations}
where
\begin{subequations}
\begin{align}
&\breve{z}\left( t \right) \!\triangleq\! \left[
    \begin{array}{c;{1pt/1pt}c;{1pt/1pt}c;{1pt/1pt}c;{1pt/1pt}c;{1pt/1pt}c}
        \!\!\!{{\breve{x}_1}\left( t \right)} \!\!&\! {\ldots} \!\!&\! {\breve{x}_{\left| \mathbb{V} \right|}}\left( t \right) \!\!&\! {\breve{v}_1}\left( t \right) \!\!&\! {\ldots} \!\!&\! {\breve{v}_{\left| \mathbb{V} \right|}}\left( t \right)
    \end{array}\!\!\!
\right]^\top\!,  \label{eq:ssd1}\\
&A_{\sigma(t)} \!\triangleq\! \left[
    \begin{array}{c;{2pt/2pt}c}
        \mathbf{0}_{\left| \mathbb{V} \right| \times \left| \mathbb{V} \right|} & \mathbf{1}_{\left| \mathbb{V} \right| \times \left| \mathbb{V} \right|}\\ \hdashline[2pt/2pt]
        -\mathcal{L}_{\sigma(t)} & -\mathbf{1}_{\left| \mathbb{V} \right| \times \left| \mathbb{V} \right|}
    \end{array}
\right],\label{eq:nm0} \\
&C \!\triangleq\! \left[
    \begin{array}{c;{1pt/1pt}c}
       C_{1} & C_{2}
          \end{array}
\right],\label{eq:nmok}\\
&{C_j} \!\triangleq\! \left[
    \begin{array}{c;{1pt/1pt}c}
   \!\!\!\emph{\emph{diag}}\!\left\{ c_{1j}, \ldots , c_{\left| \mathbb{M} \right|j} \right\} \!\!\!&\!\! \mathbf{0}_{\left| \mathbb{M} \right| \times \left( {\left| \mathbb{V} \right| - \left| \mathbb{M} \right|} \right)}\!\!\!\!
    \end{array}\right]\!, j \!=\! 1, 2\label{eq:u1b}\\
    &D \!\triangleq \left[
    \begin{array}{c;{1pt/1pt}c;{1pt/1pt}c}
   \!\!\mathbf{0}_{\left| \mathbb{M} \right| \times {\left| \mathbb{V} \right|} } \!\!\!&\!\! \emph{\emph{diag}}\!\left\{ d_{1}, \ldots , d_{\left| \mathbb{M} \right|} \right\}  \!\!\!&\!\! \mathbf{0}_{\left| \mathbb{M} \right| \times {\left( {\left| \mathbb{V} \right| - \left| \mathbb{M} \right|} \right)}} \!\!\!\!
    \end{array}\right]\!,\label{eq:nmokbb}\\
    \!\!\!\!\!&\breve{g}(t) \!\triangleq\! \left[\!\!
    \begin{array}{c;{1pt/1pt}c}
        {\bf{0}}_{\left| \mathbb{V} \right|}^\top & \bar{g}^{\top}(t)
    \end{array}
\!\!\right]^\top,\label{eq:nm2}\\
&[{\bar{g}}\left( t \right)]_{i} \!\triangleq\! \left\{ \begin{array}{l}
\hspace{-0.2cm}{\breve{g}_i}\left( t \right),i \in \mathbb{K}\\
\hspace{-0.2cm}0, \hspace{0.62cm}i \in {\mathbb{V}} \backslash \mathbb{K}.
\end{array} \right.\label{eq:u1}
\end{align}
\end{subequations}In addition, we consider the system \eqref{eq:s1} in the absence of attacks:
\begin{subequations}
\begin{align}
\dot{{{{z}}}}\left( t \right) &= A_{{{\sigma}}(t)}{{{z}}}\left( t \right),\label{eq:s21}\\
y\left( t \right) &= C{z}(t).\label{eq:s22}
\end{align}\label{eq:s2}\end{subequations}

\section{Problem Formulation}
We make the following assumptions on the attacker and the defender.
\begin{asm}
The attacker
\begin{itemize}
\item is aware that the changes in system dynamics are used by the defender (system operator);
\item knows the output matrix, the initial topology and the switching times before the first topology switching;
\item needs a non-negligible time to compute and update the attack policy and identify the newly activated topology.
\item can record the newly obtained knowledge of network topology into her memory.
\end{itemize}
\label{thm:att}
\end{asm}
\begin{asm}
 The defender
\begin{itemize}
  \item designs the switching times (when to switch) and the  switching topologies (what topology to switch to);
  \item chooses candidate agents to monitor, i.e., the monitoring agent set $\mathbb{M}$, for attack detection;
  \item knows that the states of~(\ref{eq:ooiio}) in the absence of attack are continuous with respect to time, i.e., $x\left( {{t^ - }} \right) = x\left( t \right) = x\left( {{t^ + }} \right)$ and $v\left( {{t^- }} \right) = v\left( t \right) = v\left( {{t^ + }} \right)$;
  \item  has no knowledge of the attack starting, pausing and resuming times, and the misbehaving agents.
\end{itemize} \label{thm:auatt}
\end{asm}

\subsection{Attack-Starting Time}
We now use the following to describe the system \eqref{eq:sl2} in the absence of attacks:
\begin{subequations}
\begin{align}
\dot{z}\left( t \right) = A{z}\left( t \right),\\
y(t) = C{z}\left( t \right).
\end{align}\label{eq:sl1}\end{subequations}
Let us first recall the properties of ZDA to review the prior defense strategies. \begin{lem}~\cite{teixeira2012attack}
Consider the systems \eqref{eq:sl2} and \eqref{eq:sl1}. Under the ZDA policy \eqref{eq:czeo}, the outputs and system states satisfy
\begin{align}
y\left( t \right) &= \breve{y}\left( t \right), t \ge 0 \label{eq:sp}\\
\breve{z}\left( t \right) &= z\left( t \right) + \mathbf{z}_{0}{e^{\eta  {t } }}.\label{eq:op}
\end{align}
\end{lem}

If the attack-starting time, denoted by $\kappa$, is not the initial time, the ZDA signal in \eqref{eq:sl2} and the state \eqref{eq:op} would intuitively update as
\begin{align}
{\breve{g}}(t) &=  \begin{cases}
g{e^{\eta (t - \kappa )}},&t \ge \kappa \\
\bf{0}_{\bar{o}},&\text{otherwise},
 \end{cases}.\label{zdanon}\\
\breve{z}\left( t \right) &=  \begin{cases}
z\left( t \right),&t \in  [0, \kappa) \\
z\left( t \right) + \mathbf{z}_{0}{e^{\eta  {(t-\kappa)} }},&t \in  [\kappa, \infty).
 \end{cases}.\label{statenew}
\end{align}
We note that the system \eqref{eq:sl2} is not subject to attack when $t < \kappa$, which results in the first term in \eqref{statenew}. Consequently, $\breve{z}\left( \kappa \right) =  z\left( \kappa^{-} \right)$. Moreover, after the attacker launches attack at $t = \kappa$, according to the second term in \eqref{statenew} we have $\breve{z}\left( \kappa \right) = z\left( \kappa \right) + \mathbf{z}_{0}$. Since $\mathbf{z}_{0}$ is a non-zero vector, we have $\breve{z}\left( \kappa \right) \ne z\left( \kappa \right) = \breve{z}\left( \kappa^{-} \right)$, i.e., if $\kappa \ne 0$, the attack causes ``jump" of the system state $\breve{z}\left( \kappa \right)$, which contradicts with the continuity of $\breve{z}\left( \cdot \right)$ w.r.t. time. Therefore, we conclude here that the defense strategy against ZDA -- according to Definition \ref{defzda} -- implicitly assumes that the attack-starting time is the initial time.

\subsection{Naive Attacker}
 Strategically changing system dynamics has been demonstrated to be an effective approach to detect system-based stealthy attacks, see e.g., ZDA \cite{T12} and $\mathcal{C}_{k}$/$\mathcal{C}$ stealthy attacks \cite{teixeira2014security}. The core idea behind this defense strategy is the intentional mismatch between the models of the attacker and the defender.  Specifically, the attacker uses the original system dynamics to make the stealthy attack decision before the system starts to operate, while the defender strategically changes the system dynamics at some operating point in time.
 However, this defense strategy assumes that the attacker has no capability of inferring the altered system dynamics.  Otherwise, the attacker can make a synchronous attack decision according to the inferred changed dynamics to evade the detection.

To remove these unrealistic assumptions, \cite{mao2017strategic} first investigated the attack policy for the attack signal \eqref{zdanon}, subject to the properties \eqref{eq:sp} and \eqref{statenew}. A corresponding  defense strategy of topology switching was then proposed to detect this ZDA variation in the multi-agent systems like \eqref{eq:ooiio}. However, the attack signal \eqref{zdanon}  in \cite{mao2017strategic} assumes that once the attacker launches the attack, she never changes the attack strategy.
This further indicates the attacker's assumption that the defender follows the defense strategy proposed in \cite{T12,teixeira2014security} to switch the topology only several times after the system starts the operation. In other words, the attacker assumes that
 without the knowledge of the attack-starting time the defender cannot switch the topology infinitely many times over infinite time to detect her.  Otherwise, the attacker can try to infer the topologies online and update the stealthy attack decision accordingly.

We note that the recently developed inference algorithms can use the data of the agents' states in a \emph{non-negligible} time interval to exactly infer either undirected \cite{van2019topology} or directed network topology \cite{mao2019inference1,mao2019inference2}. If the attacker has such inference ability and is aware of the topology-switching defense strategy, as made in Assumption \ref{thm:att}, she can avoid detection via the following strategy:
\begin{itemize}
  \item pauses the attack before the incoming switching times;
  \item resumes, and if necessary, updates the attack, after the newly switched/activated topology is inferred.
\end{itemize}

In this paper, we first systematically study the behavior of such  realistic ZDA variation that is referred to as intermittent ZDA. We then investigate its detectability. The  two problems are stated formally as follows:

\textbf{Problem I}: Following which attack policy, the attacker can make stealthy attack decision?

\textbf{Problem II}: Following which defense strategy, the defender can detect intermittent ZDA?

\section{Problem I: Attack Policy}
For convenience, we refer to $\mathbb{T}$ as the set of topologies under which the attacker injects attack signals to control inputs, and we refer to ${\xi_{k}}$ and $\zeta_{k}$ as the attack-resuming and attack-pausing times over the active topology intervals $[t_{k}, t_{k+1})$, $k \in \mathbb{N}_{0}$, respectively.

The ZDA signals injected into the control input and the monitored output of the system~(\ref{eq:oofn}) with intermittent pausing and resuming behaviors are described as:
\begin{align}
 \!\! \!{\breve{g}_i}\!\left( t \right)  = \left\{ \begin{array}{l}
\hspace{-0.2cm}{g^{\sigma( \!t_{k} \!)}_{i}} \!{e^{\eta_{\sigma(t_{k})}({t - {\xi_{k}}})}} \!, \hspace{0.4cm}t  \!\in\!  \left[ {{\xi_{k}},{{\zeta_{k}}}} \right) \!\subseteq\! \left[{{t_k},{t_{k + 1}}} \!\right) \\
\hspace{-0.2cm}0,\hspace{2.5cm}\text{otherwise}.
\end{array} \right ..\label{eq:ads}
\end{align}

To analyze this ZDA, we review the monitored output~(\ref{eq:oox3}) at the first ``pausing" time $\zeta_{0}$:
\begin{align}
{\breve{y}_i}\left( \zeta^{-}_{0} \right) = c_{i1}\breve x_{i}(\zeta^{-}_{0}) + c_{i2}\breve v_{i}(\zeta^{-}_{0}) + d_i{\breve{g}}_i\left( \zeta^{-}_{0} \right), \forall i \!\in\! \mathbb{M},\nonumber
\end{align}
which implies that ${\breve{y}_i}\left( \zeta^{-}_{0} \right) = {\breve{y}_i}\left( \zeta_{0} \right)$ if and only if $d_i{\breve{g}_i}\left( \zeta^{-}_{0} \right) = d_i{\breve{g}_i}\left( \zeta_{0} \right)$, since ${\breve{v}_i}\left( \zeta^{-}_{0} \right) = {\breve{v}_i}\left( \zeta_{0} \right)$ and ${\breve{x}_i}\left( \zeta^{-}_{0} \right) = {\breve{x}_i}\left( \zeta_{0} \right)$. Meanwhile, the defender knows that the velocity and position states are always continuous with respect to time, and hence the monitored outputs must be continuous as well. Therefore, to avoid the ``jump" on monitored outputs to maintain the stealthy property~\eqref{stealthy},  the attacker cannot completely pause the attack, i.e., whenever the attacker pauses injecting ZDA signals to control inputs at pausing time $\zeta_{k}$, she must continue to inject the same attack signals to monitored outputs~(\ref{eq:oox3}):
\begin{align}
{\breve{y}}\left( t \right) = C{\breve{z}}\left( t \right) + D\sum\limits_{m = 0}^{k} {\breve{g}}\left( \zeta^{-}_{m} \right),t \in \left[ {\zeta_{k}, {\xi_{k + 1}}} \right).\label{eq:ao2}
\end{align}

Based on the above analysis, for the ZDA policy consisting of ``pausing attack" and ``resuming attack" behaviors to remain stealthy, it should satisfy (\ref{eq:ao2}) and
\begin{subequations}
\begin{align}
\mathbf{z}\left( {{t_0}} \right) &\in  \widehat{\mathbf{N}}^{k}_{1} \bigcap \widetilde{\mathbf{N}}^{k}_{1}, \label{eq:ap1}\\
\left[
    \begin{array}{c}
        \mathbf{z}\left( {{{\xi _k}}} \right)  \\ \hdashline[2pt/2pt]
        -\breve{g}\left( {{{\xi_k}}} \right)
    \end{array}
\right] &\in \ker \left( {{\mathcal{P}_r}} \right), \forall \sigma \!\left( {{{\xi_k}}} \right) \!\in\! \mathbb{T}\label{eq:ap2}
\end{align}\label{eq:zkk},
\end{subequations}
where
\begin{align}
&\widehat{\mathbf{N}}^{k}_{k} = \ker ({\mathcal{O}}_{k} ),\label{eq:cm1}\\
&\widehat{\mathbf{N}}^{k}_{q} = \ker({\mathcal{O}}_{q} ) \bigcap e^{-A_{\sigma(t_{q})}(\tau_{q} - \left( {{\zeta _q} - {{\xi_q}}} \right))}\mathbf{N}^{m}_{q+1}, 1 \leq q \leq m \!-\! 1\label{eq:cm2}\\
&\widetilde{\mathbf{N}}^{k}_{k} = \ker ({\widetilde{\mathcal{O}}}_{k} ),\label{eq:cm3}\\
&\widetilde{\mathbf{N}}^{k}_{q} = \ker({\widetilde{\mathcal{O}}}_{q} ) \bigcap e^{-A_{\sigma(t_{q})}(\tau_{q} - \left( {{\zeta _q} - {{\xi_q}}} \right))}\mathbf{N}^{m}_{q+1}, 1 \leq q \leq m \!-\! 1\label{eq:cm4}\\
&{{\mathcal{O}}}_{r} \triangleq \left[\!\!\!
    \begin{array}{c;{1pt/1pt}c;{1pt/1pt}c;{1pt/1pt}c}
        {{\left( {C} \right)^\top}} & {{\left( {C{A^{2}_r}} \right)^\top}} & \ldots & {{\left( {CA_r^{2\left| \mathbb{V} \right|-1}} \right)^\top}}
    \end{array}\!\!\!
\right]^\top,  \label{eq:om1}\\
&{\widetilde{\mathcal{O}}}_{r} \triangleq \left[\!\!\!
    \begin{array}{c;{1pt/1pt}c;{1pt/1pt}c;{1pt/1pt}c}
        {{\left( {C{A_r}} \right)^\top}} & {{\left( {C{A^{2}_r}} \right)^\top}} & \ldots & {{\left( {CA_r^{2\left| \mathbb{V} \right|}} \right)^\top}}
    \end{array}\!\!\!
\right]^\top,  \label{eq:om1poa}\\
&{\mathcal{P}_r} \triangleq  \left[
    \begin{array}{c;{2pt/2pt}c}
        {\eta_{r}{{\bf{1}}_{2\left| \mathbb{V} \right| \times 2\left| \mathbb{V} \right|}} - {A_{r}}} & {\bf{1}}_{2\left| \mathbb{V} \right| \times 2\left| \mathbb{V} \right|}\\ \hdashline[2pt/2pt]
        - C  & D
    \end{array}
\!\right], \label{eq:pal21} \\
&{\mathbf{z}}= \left[
    \begin{array}{c;{1pt/1pt}c}
        \!\!{\mathbf{x}^\top} \!\!\!&\! {\mathbf{v}^\top}\!\!\!\!
    \end{array}
\right]^\top \!\triangleq \! \breve{z} - z \!= \!\left[
    \begin{array}{c;{1pt/1pt}c}
        \!\!{\breve{x}^\top} - {x^\top} \!\!\!&\! {\breve{v}^\top} - {v^\top}\!\!\!
    \end{array}
\right]^\top\!\!\!\!,\label{eq:ed0}
\end{align}
with $\tau_{q}$ denoting the dwell time of the activated topology, i.e., $\tau_{q} = t_{q} - t_{q-1}$.

\begin{rem}
By the proof of Theorem in \cite{tanwani2012observability}, $\widehat{\mathbf{N}}^{k}_{q}$ is obtained via recursive computation consisting of \eqref{eq:cm1} and \eqref{eq:cm2}, which holds for $\widetilde{\mathbf{N}}^{k}_{q}$ as well. The recursive computations indicate that the attacker also needs the knowledge of inferred switched topologies recorded in her memory to make attack decision.
\end{rem}

\begin{prop} \cite{mao2019novel}
Under the stealthy attack policy consisting of~(\ref{eq:ao2}) and~(\ref{eq:zkk}), the states and monitored outputs of the systems~(\ref{eq:s2}) and~(\ref{eq:s1}) in the presence of attack signal~(\ref{eq:ads})  satisfy
\begin{align}
\breve{y}\left( t \right) &= y\left( t \right), t \in \left[ {{{t_0}},{{t_{k+1}}}} \right),\label{eq:cad}\\
\breve{z}\left( t \right) &= z\left( t \right) + {e^{\eta_{\sigma(t_{k)}}\left( {t - {{\xi_k}}} \right)}}{\mathbf{z}\left( {{{\xi_k}}} \right)}, t \in \left[ {{{\xi _k}},{{\zeta_{k}}}} \right).\label{eq:rs2as}
\end{align}\label{thm:ccoo2}
\end{prop}

\section{Problem II: Detectability}
The following theorem presents the  detectability of intermittent ZDA.
\begin{thm} \cite{mao2019novel}
Consider the system~(\ref{eq:s1}) in the presence of attack signals~(\ref{eq:ads}). Under the defense strategy:
\begin{subequations}\label{eq:MA_sys}
\begin{align}
\mathcal{L}_{r} \hspace{0.1cm}\emph{\emph{has}}\hspace{0.1cm} \emph{\emph{distinct}}\hspace{0.1cm} \emph{\emph{eigenvalues}} \hspace{0.1cm} \emph{\emph{for}} \hspace{0.1cm} \forall r \in \mathbb{S}\label{eq:dfs1}\\
\exists i \in \mathbb{M}: \left[ {{Q_r}} \right]_{i,j} \ne 0, \forall j \in \mathbb{V}, \forall r \in \mathbb{S} \label{eq:dfs2}
\end{align}\label{eq:dfs}\end{subequations}
\begin{itemize}
  \item if the monitored agents output the full observations of their velocities (i.e., ${c_{i1}} = 0 \hspace{0.1cm} \emph{\emph{and}} \hspace{0.1cm} {c_{i2}} \ne 0$ for $\forall i \in \mathbb{M}$), the intermittent ZDA is detectable and \begin{align}
 \mathbf{N}^{\infty}_{1} = \left\{ {{{\mathbf{0}}_{2\left| \mathbb V \right|}}}, \left[
    \begin{array}{c;{1pt/1pt}c}
       \!\!\!\mathbf{1}_{\left| \mathbb{V} \right|}^\top \!\!\!&\!\! \mathbf{0}_{\left| \mathbb{V} \right|}^\top \!\!\!
          \end{array}
\right]^\top \right\};\label{eq:oa1}
\end{align}
  \item if the monitored agents output the full observations of their positions (i.e., ${c_{i1}} \ne 0 \hspace{0.1cm} \emph{\emph{and}} \hspace{0.1cm} {c_{i2}} = 0$ for $\forall i \in \mathbb{M}$), the intermittent ZDA is detectable but \begin{align}
\mathbf{N}^{\infty}_{1} = \left\{ {{{\mathbf{0}}_{2\left| \mathbb V \right|}}} \right\};\label{eq:oa2}
\end{align}
  \item if the monitored agents output the partial observations (i.e., ${c_{i1}} \ne 0 \hspace{0.1cm} \emph{\emph{and}} \hspace{0.1cm} {c_{i2}} \ne 0$ for $\forall i \in \mathbb{M}$), and $c_{i1} = c_{i2}, \forall i \in \mathbb{M}$, the kernel of the observability matrix satisfies
  \begin{align}
 \mathbf{N}^{\infty}_{1} = \left\{ {{{\mathbf{0}}_{2\left| \mathbb V \right|}}}, \left[
    \begin{array}{c;{1pt/1pt}c}
       \!\!\!\mathbf{1}_{\left| \mathbb{V} \right|}^\top \!\!\!&\!\! -\mathbf{1}_{\left| \mathbb{V} \right|}^\top \!\!\!
          \end{array}
\right]^\top \right\};\label{eq:oa1k}
\end{align}
and the intermittent ZDA is detectable if
\begin{align}
\xi_{0} > t_{0} \hspace{0.1cm} \emph{\emph{or}} \hspace{0.1cm} D = {\mathbf{0}_{\left| \mathbb{M} \right| \times 2\left| \mathbb{V} \right|}}, \label{eq:nad1}
\end{align}
where $Q_r$ is the orthogonal matrix of $\mathcal{L}_{r}$, and  ${\mathbf{N}}^\infty_{1}$ is recursively computed by
\begin{subequations}
\begin{align}
\!\!\!\!\mathbf{N}^{m}_{m} &\!=\! \ker\left ( {\mathcal{O}}_{m} \right )\\
\!\!\!\!\mathbf{N}^{m}_{q} &\!=\! \ker({\mathcal{O}}_{q} ) \bigcap e^{-A_{\sigma(t_{q})}\tau_{q}}\mathbf{N}^{m}_{q+1}, 1 \!\leq\! q \!\leq\! m \!-\! 1
\end{align}\label{eq:kerc}\end{subequations}
with ${\mathcal{O}}_{q}$ given by~(\ref{eq:om1}).
\end{itemize}
\label{thm:thd}
\end{thm}

\begin{rem}
Under the defense strategy~(\ref{eq:dfs}), the result \eqref{eq:oa2} means  if the monitored agents output full observations of positions, the system~(\ref{eq:s2}) is observable at any time $t > t_{0}$. As a result, using the available data of sensor outputs \eqref{eq:s22}, the attacker can infer the global system state and the global initial condition. While the results~(\ref{eq:oa1}) and~(\ref{eq:oa1k}) show that if the monitored agents output full observations of velocity or partial observations, the privacy of full states of non-monitored agents are preserved, which would be useful in defending ZDA in cooperation with topology attack, since the stealthy topology needs the data of real-time states to decide target links to the attack \cite{mao2019novel}. Therefore, for the purpose of privacy preserving of non-monitored agents' states, consequently, restricting the scope of target links of the stealthy topology attack, the defender (system operator) has to abandon full observation of the position.
\end{rem}

\begin{rem}
The compact defense strategy also includes a strategy on switching times. The building block of our defense strategy is the time-dependent topology switching that does not need a central unit to trigger links to switch. The critical reason that we do not consider state-dependent topology switching is that the attack signals injected into control input may generate Zeno behaviors~\cite{ames2005sufficient}, such that the control protocol~(\ref{eq:lci}) becomes infeasible. For this part of work, we refer the readers to \cite{mao2019novel,mao1}.
\end{rem}

\section{Simulation}
\begin{figure}[http]
\centering{
\includegraphics[scale=0.40]{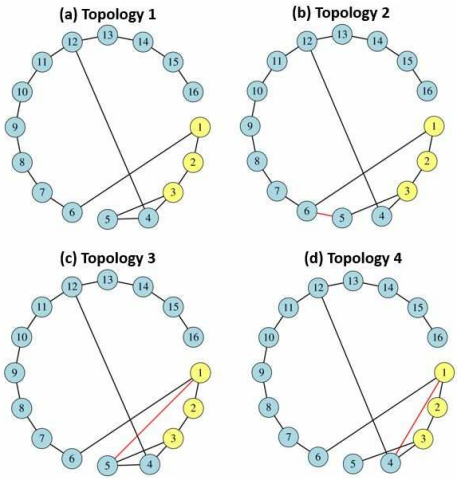}
}
\caption{Network topologies for intermittent ZDA, where agents 1, 2 and 3 are monitored agents.}
\label{fig:tpi}
\end{figure}

We consider a system with $n = 16$ agents. The initial position and velocity conditions are chosen randomly as ${x}(t_{0}) = {\left[ {2,2,2,2,2,2,2,2,4,4,4,4,4,4,4,4} \right]^ \top }$ and ${v}(t_{0}) = {\left[ {6,6,6,6,6,6,6,6,8,8,8,8,8,8,8,8} \right]^ \top }$. The considered network topologies are given in  Fig. \ref{fig:tpi}, where the agents 1, 2 and 3 are the monitored agents, and the coupling weights are uniformly set as ones. We denote $r_{i}(t) = \breve{y}_{i}\left( t \right) - {y}_{i}\left( t \right), i \in \mathbb{M} = \{1,2,3\}$, as the attack-detection signals.

We first consider the periodic topology switching sequence:  $1 \rightarrow 2 \rightarrow 1 \rightarrow 2 \rightarrow \ldots$ with the dwell times $\tau_{1} = \tau_{2} = 2$. It can be verified that neither Topology 1 nor 2 in Fig.~\ref{fig:tpi} satisfies the defense strategy \eqref{eq:dfs}. Therefore, the attacker can design undetectable intermittent ZDA by:
\begin{itemize}
  \item inject false data $\mathbf{z}(t_{0}) = \left[ {0,0,0,-1,1,0,0,0,0,0,0,0,} \right.\\
  \left. {0,0,0,0, 0,0,0,-0.5,0.5,0,0,0,0,0,0,0,0,0,0,0} \right]^\top$to the data of initial condition;
  \item inject ZDA signals ${\breve{g}_4}\left( t \right) = -1.75{e^{0.5t}}$ and ${\breve{g}_5}\left( t \right) = 1.75{e^{0.5t}}$ to the local control inputs of agents 4 and 5 for the initial Topology 1;
  \item pause the ZDA before the incoming new topology, e.g. Topology 2, since the new topology has not been inferred yet; 
  \item update the attack policy if necessary and resume the feasible attack after switching to the new topology finishes;
  \item iterate the last two steps.
\end{itemize}

\begin{figure}[http]
\centering{
\includegraphics[scale=0.35]{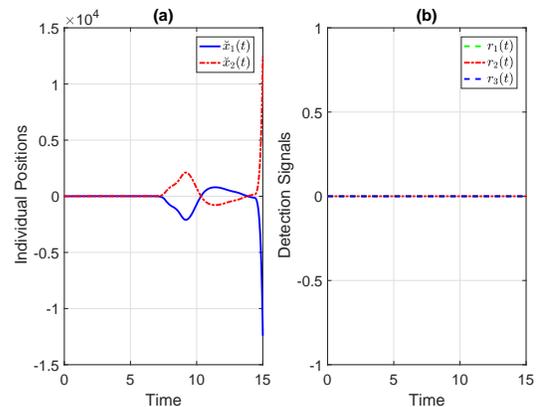}}
\caption{Trajectories of individual positions and attack-detection signals: undetectable attack under switching Topologies 1 and 2.}
\label{fig:stiz}
\end{figure}

The trajectories of some agents' positions and the attack-detection signals in Fig.~\ref{fig:stiz} show that the designed intermittent ZDA is not detected, and the stealthy attack destabilizes the system.

\begin{figure}[http]
\centering{
\includegraphics[scale=0.35]{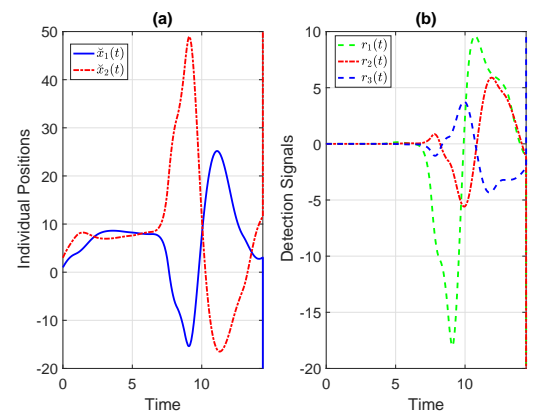}}
\caption{Trajectories of individual positions and attack-detection signals: detectable attack under switching Topologies 3 and 4.}
\label{fig:stizs}
\end{figure}
It is straightforward to show that both Topologies 3 and 4 in Fig.~\ref{fig:tpi} satisfy the defense strategy \eqref{eq:dfs}. Hence, we can turn to the following periodic topology switching sequence at some time to detect the stealthy attack: $3 \rightarrow 4 \rightarrow 3 \rightarrow 4 \rightarrow \ldots$ with the dwell times $\tau_{3} = \tau_{4} = 2$. Under the periodic sequence, the trajectories of attack-detection signals in Fig.~\ref{fig:stizs} show that the stealthy intermittent ZDA is successfully detected.

\section{Conclusion}
This paper introduces one ZDA variant for a scenario where the attacker is informed about the switching strategy of the defender: intermittent ZDA where the attacker pauses, and updates and resumes ZDA in conjunction with the knowledge of  switching topology  and dwell times. A defense strategy without requiring any knowledge of the set of misbehaving agents or the start, pause and resume times of the attack is proposed to detect the intermittent ZDA.

\section{Acknowledgment}
This work has been supported in part by NSF (award
numbers CMMI-1663460 and ECCS-1739732), and Binghamton University--SUNY, Center for Collective Dynamics of Complex Systems ORC grant.

\bibliographystyle{IEEEtran}
\bibliography{refII}
\end{document}